\documentclass[preprint,showpacs,preprintnumbers,amsmath,amssymb]{revtex4}
\usepackage{graphicx}
\usepackage{natbib}
\begin{document}


\title{Equivalence of Covariant and Light Front QED: Generating Instantaneous Diagrams \\}

\author{Swati M. Patel}
 
\author{Anuradha Misra }
\email{misra@physics.mu.ac.in}
\affiliation{Department of Physics\\
University of Mumbai,\\
Santa Cruz(E), Mumbai,\\
India-400098}%

\date{\today}
\begin{abstract}
One loop 
 expressions for fermion self energy, vacuum polarization and vertex 
correction in light-front time ordered 
perturbation theory(LFTOPT) can be obtained from respective   covariant expressions by performing $k^-$ integration.  In 
an earlier work, we have  shown that 
the the third term in the doubly transverse gauge propagator is necessary to 
generate the diagrams involving instantaneous photon exchange both in case of 
fermion self energy as well as vertex correction. In this work,we show that instantaneous photon exchange
diagrams in fermion self-energy as well as the IR singular terms in the propagating diagrams 
 can be generated by taking the asymptotic limit of the covariant expression, if one 
uses the commonly used two term photon propagator.  We also show that this method reproduces the IR singular terms in propagating diagrams of vacuum polarization.

\end{abstract}

\pacs{11.15.Bt,12.20.-m}
\maketitle

\section{\label{sec:level1}Introduction}

The issue of equivalence of covariant perturbation theory and light 
front Hamiltonian perturbation theory has attracted a lot of 
attention in recent years \citep{bakker94,schoon98,paston01,bmcy05}. 
It is important to establish equivalence between the two approaches 
as light front field theory has "spurious" divergences not present 
in covariant perturbation theory and it is necessary to understand 
how these are generated in order to establish a correspondence 
between light front expressions and the covariant expressions. One 
of the approaches consists of  establishing equivalence at the 
Feynman diagram level wherein the covariant expression for a Feynman 
diagram is integrated over the light cone energy, $k^-$, to generate 
all the diagrams of light front perturbation theory \citep{bakker94}. 
Bakker {\it etal} \citep{bakker94} have given a general algorithm 
for proving equivalence in theories involving scalars as well spin 
-$\frac{1}{2}$ particles. Equivalence at Feynman  diagram level in 
Yukawa theory has been discussed in detail in \citep{schoon98}. 
Correspondence between light-front Hamiltonian approach and the 
Lorentz-covariant approach has been discussed for QED $1+1$ and also 
for QCD by bosonization of the model \citep{paston01}. 

As far as $3+1$ dimensional theories are concerned, equivalence of 
LFQED and covariant QED in Coulomb gauge has been proved within the 
framework of Feynman-Dyson-Schwinger theory \citep{rohrlich74}. 
However, not much work  has been done on proving equivalence for QED 
at the Feynman diagram level. In a previous work \citep{amswa05} we 
had addressed the issue of equivalence of light-front QED, 
\citep{wilson91} and covariant QED at the Feynman diagram level. In 
\citep{amswa05},  we have shown how one can obtain all the 
propagating as well as instantaneous diagrams by performing the 
$k^-$ - integration carefully. The feature that sets  QED apart from 
other cases considered in literature is the presence of diagrams 
involving instantaneous photon exchange. Our previous study was 
aimed at generating these expressions in the  diagram based approach.  
It was shown that the equivalence cannot be established by performing 
$k^-$ integration if one uses the commonly used two term photon 
propagator in light cone gauge, \citep{wilson91,hari98}:
\begin{equation}
d_{\mu\nu} = \frac{1}{k^2+i\epsilon}\bigg[-g_{\mu\nu} + 
\frac{\delta_{\mu+}k_{\nu}+\delta_{\nu+}k_{\mu}}{k^+}\bigg]
\label{eq:prop1}
\end{equation}  
 
However, if one uses the three term  photon propagator term 
\citep{hari98,yan73,rohrlich74,brodsky01,suzuki03,suzuki04} given by  

\begin{equation}
d_{\mu\nu} = \frac{1}{k^2+i\epsilon}\bigg[-g_{\mu\nu} + 
\frac{\delta_{\mu+}k_{\nu}+\delta_{\nu+}k_{\mu}}{k^+}-
\frac{k^2\delta_{\mu+}\delta_{\nu+}}{(k^+)^2}\bigg]
\label{eq:prop2}
\end{equation}       
    
then one can generate the diagrams involving instantaneous photon 
exchange also which completes the proof of equivalence. In present 
work, we give an alternative proof of  equivalence of covariant and 
LFQED at the Feynman diagram level using the two term photon 
propagator. We will show that one can  use the asymptotic method  
proposed by Bakker {\it etal} \citep{bmcy05} to generate the 
instantaneous photon exchange diagrams  for one loop self energy 
correction. This method does not require the third term in the 
photon propagator. There has been some debate in literature 
over the relevance of the third term in the gauge boson propagator. 
It is usually dropped on the grounds that it does not propagate any 
information. In our previous work, we emphasized the importance of 
this term in proving equivalence at one loop level. The present work 
gives an alternative proof i.e. one without the need of the third 
term. However, it should not be considered as undermining the 
importance of this term. On the contrary, the present method, being 
an alternative to the three term propagator method, may be able to 
throw some light on the physical significance of this term. 

The plan of the paper is as follows: In Section ~\ref{sec:pertur}, we 
summarize  the one loop renormalization of LFQED \citep{wilson91} and 
briefly review the work of \citep{amswa05} for completeness. Here, we 
present only those results of \citep{wilson91} and \citep{amswa05} 
which are needed for our discussion. In Section ~\ref{sec:sec3}, we 
consider self energy diagram and use the asymptotic method to generate  
graph involving instantaneous photon exchange. We will show that in a 
certain asymptotic limit, the covariant expression for fermion self 
energy reduces to a sum of expression for the instantaneous photon 
exchange graph and the IR singular terms of the propagating graph. 
We also carry out a similar analysis for vacuum polarization. Since 
vacuum polarization does not have any contribution from instantaneous 
photon exchange vertex at one loop level, in this case the above 
mentioned limit reproduces only the IR singular terms in the 
propagating part. In Section ~\ref{sec:sec4}, we summarize and 
discuss our results. Appendix A contains the notations and 
basics. Appendix B contains some useful formulae. 

\section{\label{sec:pertur} PROOF OF EQUIVALENCE OF  COVARIANT AND 
LIGHT FRONT QED USING THE THREE TERM PHOTON PROPAGATOR}

In this section, we summarize the results of \citep{wilson91} on  one 
loop renormalization of light front QED in Hamiltonian formalism and 
recall how these results were obtained by performing $k^-$ 
integration in \citep{amswa05}. 

\subsection{\bf  ELECTRON MASS  RENORMALIZATION}

In light cone time ordered perturbation theory (LCTOPT), fermion self 
energy at  $O(e^2)$ has three contributions given by
\begin{equation}
\bar u(p,s^\prime)\Sigma_1(p)u(p,s) = 
\langle p,s^\prime \vert V_1 \frac{1}{p^--H_0}V_1 
\vert p,s\rangle
\label{eq:ma1}
\end{equation}

\begin{figure}
\begin{center}
\includegraphics*[height=5.2cm]{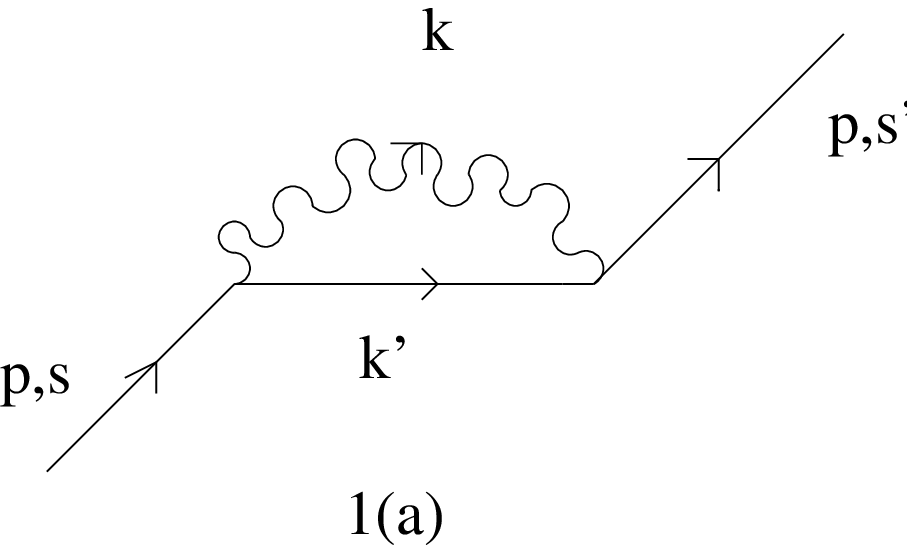}
\includegraphics*[height=5.2cm]{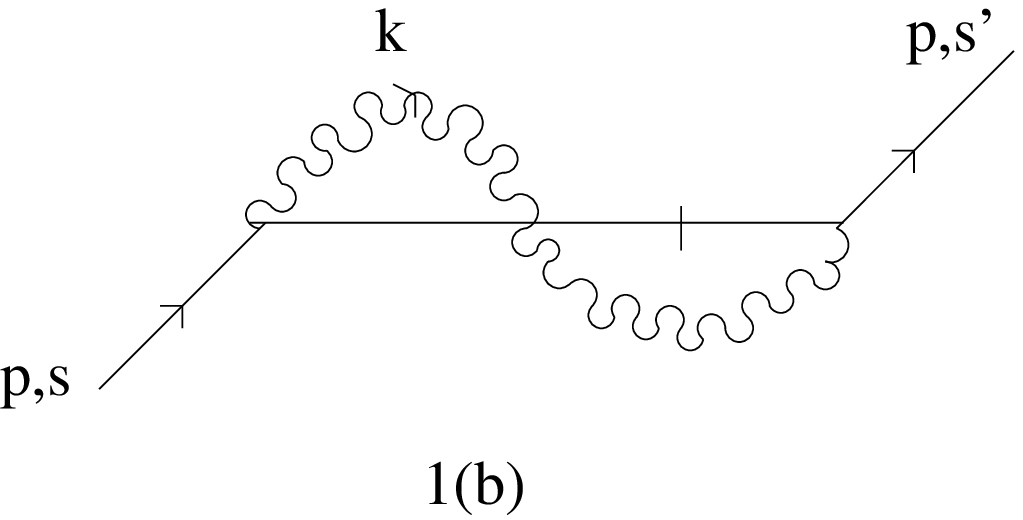}
\includegraphics*[height=5.2cm]{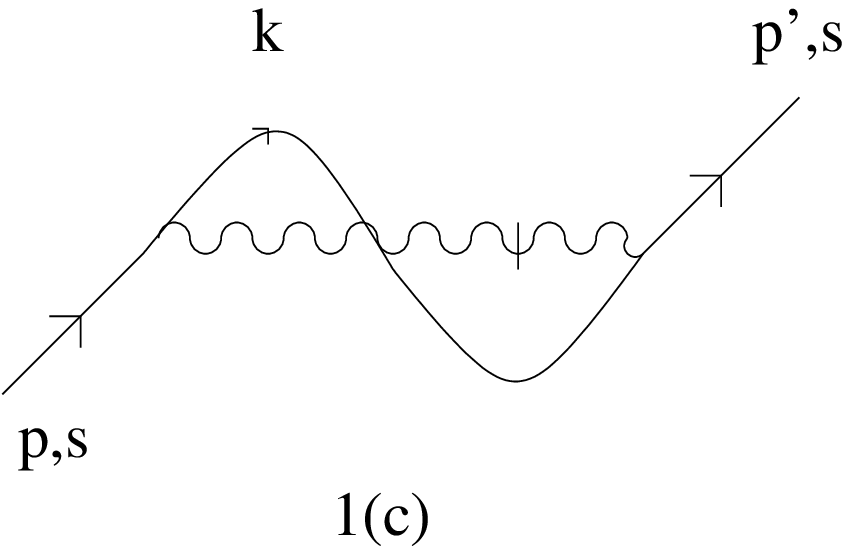}
\includegraphics*[height=5.2cm]{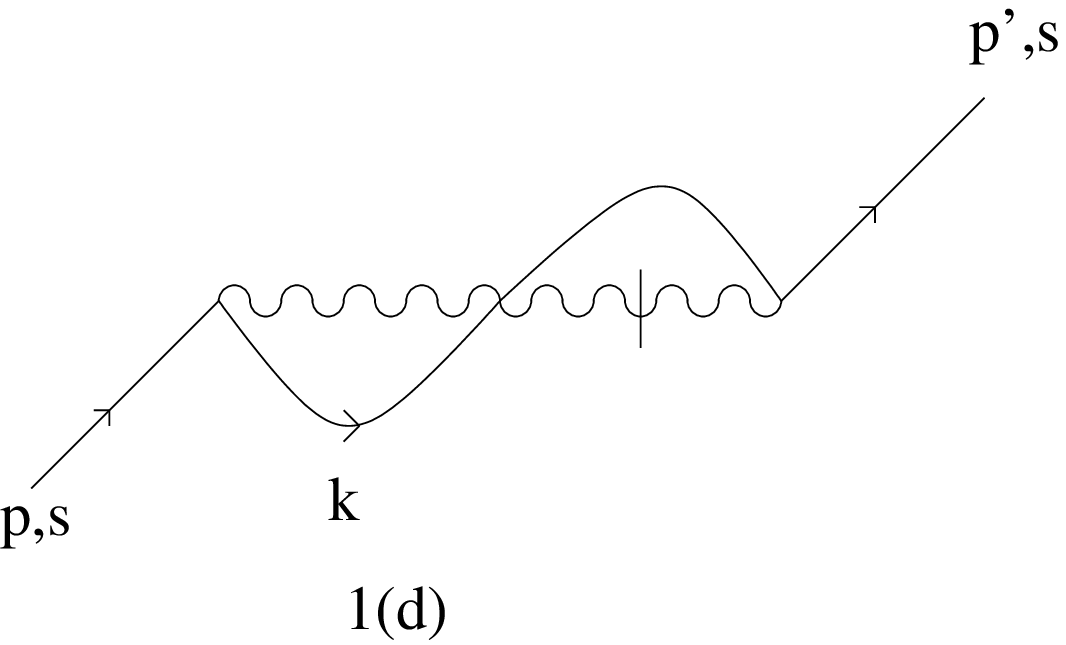}
\end{center}

\nonumber
\caption{\label{Fig1}Diagrams for electron mass shift in LFQED}
\end{figure}
corresponding to the diagram  in Fig.~\ref{Fig1}(a), 

\begin{equation}
\bar u(p,s^\prime)\Sigma_2(p)u(p,s) = 
                   \langle p,s^\prime \vert V_2 \vert p,s \rangle
\end{equation}
corresponding to diagram in Fig.~\ref{Fig1}(b) and 
\begin{equation}
\bar u(p,s^\prime)\Sigma_3(p)u(p,s) = 
                   \langle p,s^\prime \vert V_3 \vert p,s \rangle
\end{equation}
corresponding to sum of diagrams in Fig.\ref{Fig1}(c) and 
Fig.~\ref{Fig1}(d), $V_1$ is the standard three point QED vertex and 
$V_2$ and $V_3$ are $O(e^2)$ non local four point vertices 
corresponding to exchange of instantaneous fermion and photon 
respectively. Expressions for $V_1$, $V_2$ and $V_3$ are given in 
Appendix A. 

The contribution of Fig.~\ref{Fig1}(a) to $\delta m$ is given by 
Eq.~(\ref{eq:ma1}) and leads to the light cone expression for 
propagating part given by 
\begin{equation}
\delta m_a \delta_{s \sigma} = \frac{e^2}{m}\int\frac{d^2 k_{\perp}}
{(4\pi)^3}\,\int_0^{p^+} \frac{d k^+}{k^+(p^+-k^+)} 
\frac{\bar u(p,\sigma) \gamma^\mu\,( {k \llap/}^{\prime} + m)
\gamma^\nu u(p,s)d_{\mu\nu}(k)}{p^--k^--k^{\prime -}}
\label{eq:m1}
\end{equation}
where all the  momenta are on shell:
\begin{equation}
p= \bigg(p^+,\frac{p_{\perp}^2 + m^2}{2p^+},p_{\perp}\bigg)
\end{equation}
\begin{equation}
k = \bigg(k^+,\frac{k^2_{\perp}}{2k^+},k_{\perp}\bigg)
\end{equation}
and
\begin{equation}
k^{\prime} = \bigg(p^+-k^+,\frac{(p_{\perp} - k_{\perp})^2 +m^2}
{2(p^+-k^+)},p_{\perp} - k_{\perp}\bigg)
\end{equation}

The contributions of Fig.~\ref{Fig1}(b)  is
\begin{equation}
\delta m_{b}\delta_{ss^\prime}
= \frac{e^2 p^+ \delta_ {ss^\prime}}{2m}
\int\frac{d^2 k_{\perp}}{(2\pi)^3} \int_0^{+\infty} \frac{d k^+}
{k^+(p^+-k^+)}
\label{eq:m2}
\end{equation}
and the sum of contributions of Fig. ~\ref{Fig1}(c) and 
Fig.~\ref{Fig1}(d) is  
\begin{equation}
\delta m_{c}\delta_{ss^\prime}=
\frac{e^2 p^+ \delta_{ss^\prime}}{2m} 
\int \frac{d^2 k_{\perp}}{(2\pi)^3} \bigg[\int_0^{+\infty}
\frac{d k^+}{(p^+-k^+)^2} - \int_0^{+\infty}
\frac{d k^+}{(p^++k^+)^2} \bigg]
\label{eq:m3}
\end{equation}
These integrals have potential singularities at $k^+=0$ and 
$k^+=p^+$. To regularize them one introduces small cutoffs 
$\alpha$ $\beta$
\begin{equation}
\alpha\leq k^+\leq p^+-\beta
\end{equation}
and removes the pole at $k^+=p^+$ in $\delta m_{b}$ and 
$\delta m_{c}$ by principal value prescription. Using this procedure 
one obtains
\begin{equation}
\delta m_a=\frac{e^2}{2m}\int\frac{d^2k_\perp}{(2\pi)^3}
\bigg[\int_0^{p^+}\frac{dk^+}{k^+}\frac{m^2}{p\cdot k}-2
\bigg[\frac{p^+}{\alpha}-1\bigg]-ln\bigg[\frac{p^+}{\beta}\bigg]
\bigg]
\end{equation}
\begin{equation}
\delta m_b = \frac{e^2}{2m}\int\frac{d^2k_\perp}{(2\pi)^3}ln\bigg(\frac{p^+}{\alpha}\bigg)
\label{eq:mb}
\end{equation}
and 
\begin{equation}
\delta m_c = \frac{e^2}{m}\int\frac{d^2k_\perp}{(2\pi)^3}\bigg[\frac{p^+}{\alpha}-1\bigg]
\label{eq:mc}
\end{equation}
To establish equivalence, one starts with covariant expression for 
electron self energy in the light-front gauge ,
\begin{equation}
\sum(p)=\frac{(ie)^2}{2mi}\int\frac{d^4k}{(2\pi)^4}
\frac{\gamma^\mu(p \llap/-k \llap/+m)\gamma^\nu d
^\prime_{\mu\nu}(k)}{[(p-k)^2-m^2+i\epsilon][k^2-\mu^2+i\epsilon]}
\label{eq:sigma}
\end{equation}\\
where $\frac{d^\prime_{\mu\nu}}{k^2}$ is the photon propagator in 
light-cone gauge in covariant perturbation theory with 
$d^\prime_{\mu\nu}(k)$ given by Eq.~(\ref{eq:prop2}):
Substituting 
\begin{eqnarray}
 \nonumber
p \llap/-k \llap/+m
=\gamma^+\bigg[\bigg(\frac{(p_\perp-k_\perp)^2+m^2}{2(p^+-k^+)}\bigg)\bigg]
+\gamma^- (p^+-k^+)-\gamma_\perp(p_\perp-k_\perp) \\
+\gamma^+\bigg[p^--k^--\frac{(p_\perp-k_\perp)^2+m^2}{2(p^+-k^+)}\bigg]
\end{eqnarray}
and integrating over light cone energy $k^-$, one obtains 
\begin{equation}
\sum(p)= {\sum}_1^{(a)}(p) + {\sum}_1^{(b)}(p) +  {\sum}_2(p)
\end{equation}
where\\
\begin{equation}
{\sum}_1^{(a)}(p) = \frac{e^2}{m}\int\frac{d^2 k_{\perp}}
{(4\pi)^3}\,\int_0^{p^+} \frac{d k^+}{k^+(p^+-k^+)} \frac{
\gamma^\mu\,( {k \llap/}^{\prime} + m)\gamma^\nu d_{\mu\nu}(k)}
{p^--k^--k^{\prime -}}
\end{equation}\\
is the propagating part leading to $\delta m_a $. ${\sum}_2(p)$ 
is given by
\begin{equation}
{\sum}_2(p)=\frac{e^2}{2m}\int^\infty_0\frac{dk^+}{2k^+}
\int\frac{d^2k_\perp}{(2\pi)^3}\frac{\gamma^\mu\gamma^+\gamma^\nu  
d_{\mu\nu}(k)}{2(p^+-k^+)}
\end{equation}
and leads to $\delta m_b$, whereas $\sum_1^{(b)}$ arises from the 
third term in photon propagator and yields $\delta m_c$.

$\sum^{(a)}_1(p)$ differs from the covariant expression in that the 
fermion momentum in the loop is on-shell in the light front 
expression, i.e. 
\begin{equation}
k^{\prime}=\bigg(p^+-k^+,\frac{(p_\perp-k_\perp)^2+\ m^2}{2(p^+-k^+)},{p}_\perp-{k}_\perp\bigg)
\end{equation}
whereas in covariant expression it is off shell.

One should recall   that $\delta m_b$ arises when off shell momentum 
in covariant expression is replaced by on shell momentum. In fact in 
LFPT all diagrams involving instantaneous fermion exchange are 
obtained by the replacement
\begin{equation}
k^-\rightarrow k^-_{on}+(k^--k^-_{on})
\end{equation}
The first term here generates the LF propagating diagram and the 
second term generates the instantaneous fermion exchange diagram. 
Note that the resulting expression for $\delta m_a$ still has IR 
singular terms. We will show in section III that these IR singular 
terms and $\delta m_c$ can be obtained by taking the limit 
$k^+\rightarrow p^+$, $k^-\rightarrow\infty$ in the covariant 
expression.

\subsection{\bf PHOTON MASS  RENORMALIZATION}
   
In exactly the same manner as for electron self energy, the covariant 
expression for photon self energy can also be shown to be equivalent 
to the sum of the propagating and instantaneous diagrams of light 
front field theory by changing the off shell momenta to on shell 
momenta.

\begin{figure}
\begin{center}
\includegraphics*[height=5.0cm]{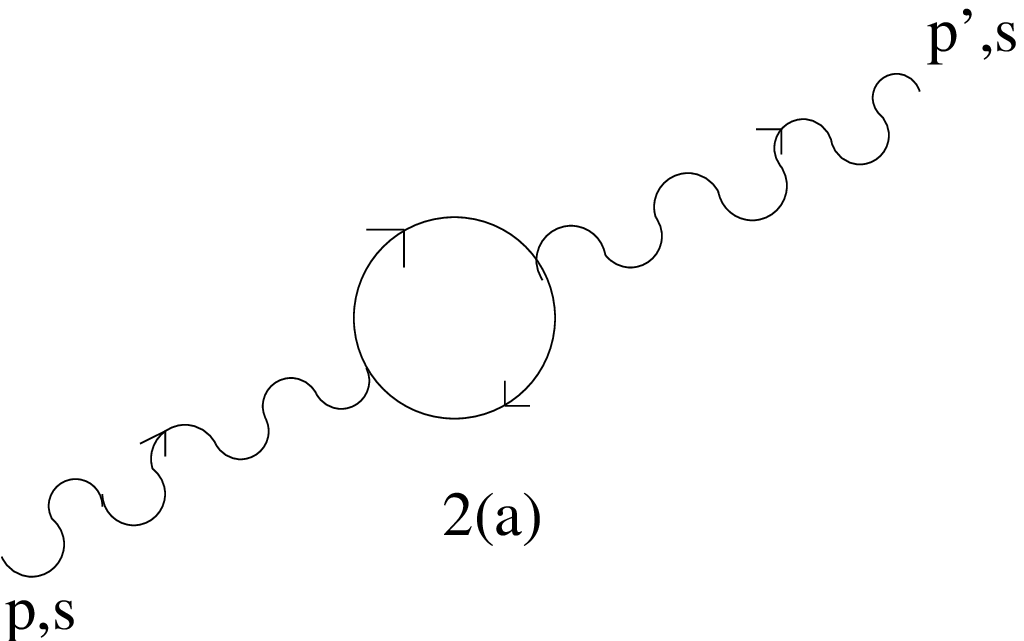}
\includegraphics*[height=5.0cm]{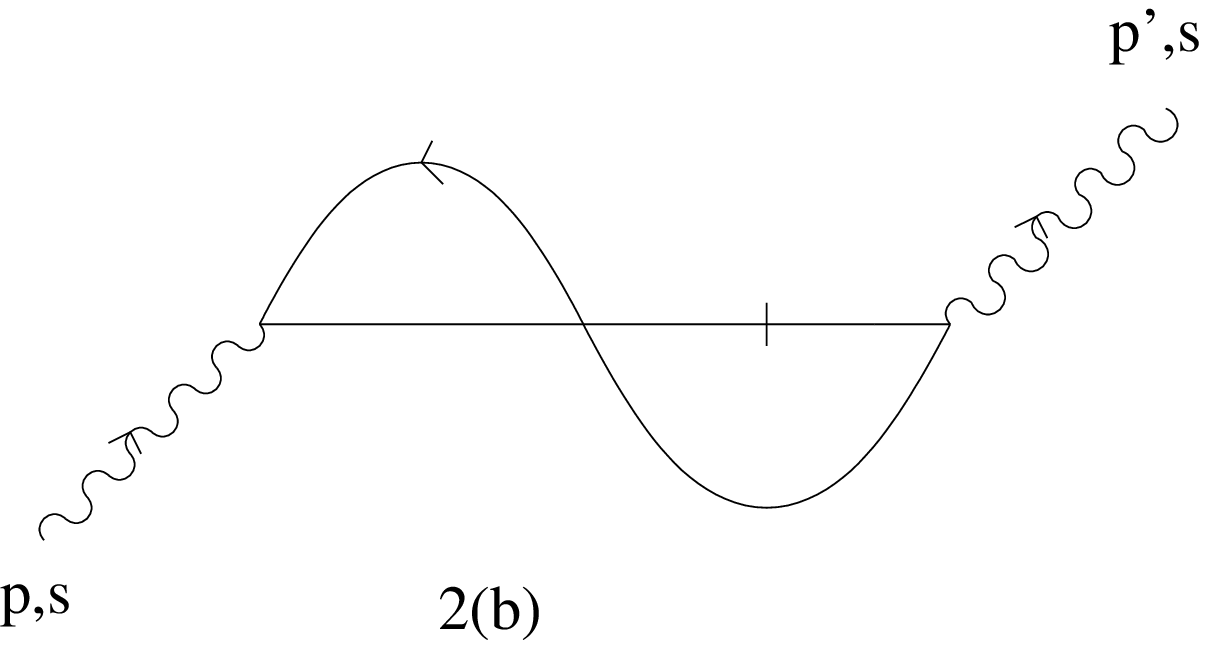}
\includegraphics*[height=5.0cm]{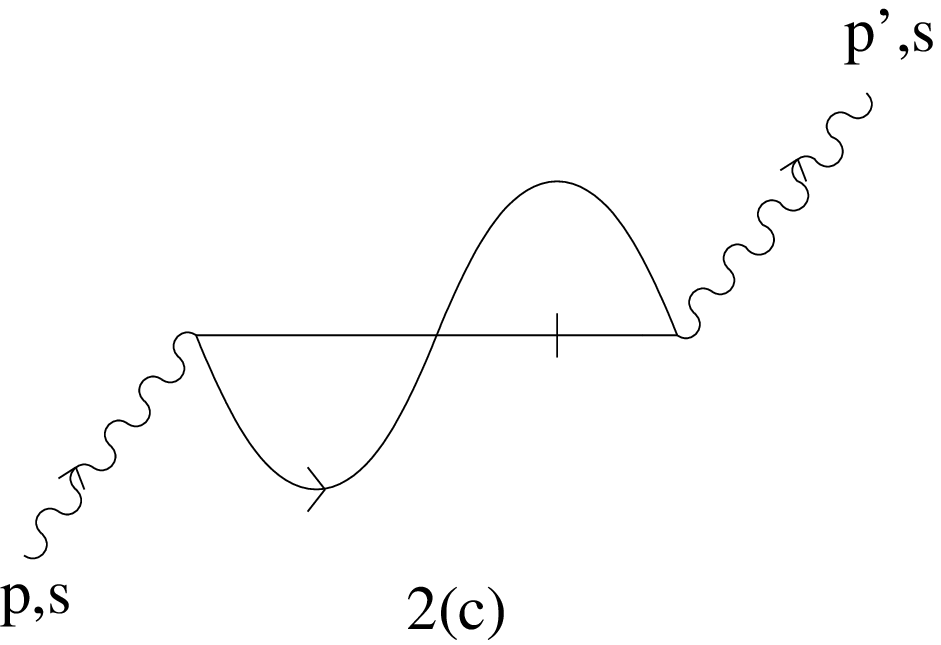}
\end{center}
\nonumber
\caption{\label{Fig2}Diagrams for vacuum polarization in LFQED}
\end{figure}

One defines a tensor $\Pi^{\mu\nu}(p)$ through
\begin{equation}
\delta\mu^2\delta_{\lambda\lambda^{\prime}}=\epsilon^\lambda_\mu(p)
\Pi^{\mu\nu}(p)\epsilon^{\lambda^\prime}_\nu(p)
\end{equation}
The corresponding diagrams are displayed in Fig.~(\ref{Fig2}) 
$\delta \mu_a^2$ is given by
\begin{equation}
\delta \mu_a^2\delta_{\lambda\lambda^{\prime}}=\left\langle p,
\lambda\left|V_1\frac{1}{p^--H_0}V_1\right|p,\lambda\right\rangle
\end{equation}

wheras the segulls are given by
\begin{equation}
\delta \mu_{b+c}^2 = \left\langle p,\lambda|V_2|p,\lambda\right\rangle
\end{equation}

Inserting appropriate sets of intermediate states and following the 
standard procedure, one obtains

\begin{equation}
\delta \mu_a^2=2e^2\int\frac{d^2k_\perp}{(4\pi)^3}
\int^{p^+-\beta}_{\alpha}\frac{dk^+}{k^+(p^+-k^+)}
\frac{tr[\epsilon\llap/^{\lambda}(p)(k\llap/+m)
\epsilon\llap/^{\lambda^\prime}(p)(k\llap/^\prime-m)]}
{p^--k^--k^{\prime-}}
\end{equation}

where,
\begin{equation}
k=\bigg[k^+,\frac{k^2_\perp+m^2}{2k^+},k_\perp\bigg]
\end{equation}

and
\begin{equation}
k^\prime=\bigg[p^+-k^+,\frac{(p_\perp-k_\perp)^2+m^2}
{2(p^+-k^+)},p_\perp-k_\perp\bigg]
\end{equation}

$\delta\mu^2$ is the sum of $\delta \mu_a^2$ and 
$\delta \mu_{b+c}^2$ where
\begin{equation}
\delta\mu^2_a=e^2\int\frac{d^2k_\perp}{(2\pi)^3}\bigg[ln\bigg[
\frac{\alpha\beta}{({p^+}^2)}\bigg]+\frac{2{k_\perp^2}}
{k_\perp^2+m^2}\bigg]
\label{eq:28}
\end{equation}
corresponds to the propagating diagram and 
\begin{equation}
\delta \mu^2_{b+c}=e^2\int\frac{d^2k_\perp}{(2\pi)^3}
\int_0^\infty dk^+\bigg[\frac{1}{p^+-k^+}-\frac{1}{p^+-k^+}\bigg]
\end{equation}
corresponds to the instantaneous fermion exchange. 

One can obtain this result from covariant expression also by 
performing the $k^-$ integration in a manner similar to the one  
sketched above for fermion self-energy diagram.

\section{\label{sec:sec3} \bf  ASYMPTOTIC METHOD AND LIGHT FRONT QED}
In this section, we will show that the diagrams involving instantaneous photon 
exchange in fermion self energy can be generated by  the 
asymptotic method discussed by Bakker {\it etal} in the context of $1+1$-dimensioal theories\cite{bmcy05}.

In general, the number of light cone energy denominators is one less 
than the number of denominators in the covariant expression. This may 
give the impression  that  one can obtain the propagating part of 
light cone expression from the covariant expression by integrating 
over light cone energy $k^-$ using the method of residues. However, 
this apparently straightforward manner of proving equivalence does 
not reproduce all the instantaneous diagrams unless one takes into 
account the contribution of arc at infinity and end point 
contributions \citep{amswa05, bmcy05}. The diagrams involving 
instantaneous fermion exchange arise in a straightforward manner 
when the fermion momenta in covariant expression are replaced by 
on shell momenta, as discussed in the previous section. We will not 
discuss this contribution here. In Ref. \citep{amswa05}, we have 
shown that the diagrams involving instantaneous photon exchange 
arise from the third term in the photon propagator of 
Eq.~(\ref{eq:prop2}). We will now show that these instantaneous 
diagrams can also be generated by taking the asymptotic limit of 
leading $k^-$ term in the covariant expression of one loop diagrams 
with conventional two term photon propagator of Eq.~(\ref{eq:prop1}). 
In addition, the IR divergent term in propagating part can also be 
generated by this method.
   The asymptotic method\cite{bmcy05} consists of isolating the divergent parts by identifying the behaviour of the integrand at asymptotic values of $k^-$ and then regularizing these divergent parts in an appropriate manner.
 In Ref.~\cite{bmcy05}, Bakker {\it etal} regularize the divergent part by 
shifting the integration variables to light-front cylindrical coordinates $k^+ = R \cos\phi$ and $k^- = R\sin\phi$. The regularized integrals are then evaluated over a finite region first (keeping R finite) and finally the limit $R \rightarrow \infty$ is taken. 

In the following, we will use the asymptotic method to isolate the divergent 
parts of one loop expressions for self energy and vacuum polarization in QED. 
We will then use the u-coordinate regularization\cite{bakker94} to evaluate these integrals.  We will show that this method reproduces the instantaneous photon exchange diagram as well as the divergent part of the propagating diagram even if one uses the two term photon propagator. We will consider the covariant expression in the limit when $k^- \rightarrow \infty$ and the light cone momentum of internal fermion line approaches zero since we are interested in generating diagrams involving instantaneous photon exchange.i.e.Figs.1(c) and (d). 

The covariant expression for electron self energy in the light-
front gauge with the two term photon propagator is  given by,
\begin{equation}
\sum(p)=\frac{(ie)^2}{2mi}\int\frac{d^4k}{(2\pi)^4}\frac{N}{D_1 D_2}
\label{eq:sigma}
\end{equation}\\
where , 
\begin{equation}
N =\gamma^\mu(p \llap/-k \llap/+m)\gamma^\nu d_{\mu\nu }(k)
\end{equation}
\begin{equation}
D_1 = k^2-\mu^2+i\epsilon
\end{equation}
\begin{equation}
D_2= (p-k)^2-m^2+i\epsilon
\end{equation}
and $\frac{d_{\mu\nu}(k)}{k^2}$ is the photon propagator in 
light-cone gauge commonly used in light front QED \citep{wilson91} 
given  by \\
\begin{equation}
d_{\mu\nu} =-g_{\mu\nu}(k)+\frac{\delta_{\mu+}k_{\nu}+
\delta_{\nu+}k_\mu}{k^+}
\end{equation}\\
Using Eq.(A.8) -(A.12), the numerator reduces to 

\begin{equation}
N = (p^+-k^+)\bigg[2\gamma^- + \frac{4\gamma^+k^-}{k^+}-
\frac{2\gamma_\perp\cdot k_\perp}{k^+}\bigg]+
(p^--k^-)2\gamma^+-\frac{2\gamma^+}{k^+}k^i(p^i-k^i) -2m
\end{equation}

Numerator in $\bar u\sum(p) u $ is obtained from this by using 
Eq(A.14) and Eq(A.15). In the limit $k^-\rightarrow\infty$, 
numerator in $\bar u\sum(p) u $ is reduced to
\begin{equation}
N^\prime = \frac{8p^+}{k^+}(p^+-k^+)k^- -4p^+k^-+ 
\frac{4p^+}{k^+}k_\perp^2 
\end{equation}

In the limit $ k^- \rightarrow \infty, k^+ \rightarrow p^+$,\\
$D_1 \rightarrow 2k^+k^-$
\\
and $\delta m $ reduces to 
\begin{equation}
\delta m_{asy}= \delta m_1 + \delta m_2 +\delta m_3
\end{equation}
where
\begin{equation}
\delta m_1  = \frac{ie^2p^+}{m}\int\frac{d^2k_\perp}{(2\pi)^4}
\int\frac{dk^+}{k^{+2}}\int\frac{dk^-}{p^--k^- - 
\frac{(p_\perp-k_\perp)^2+m^2-i\epsilon}{2(p^+-k^+)}}
\end{equation}

\begin{equation}
\delta m_2 = -\frac{ie^2p^+}{m}\int\frac{d^2k_\perp}{(2\pi)^4}
\int\frac{dk^+}{2k^+(p^+-k^+)}\int\frac{dk^-}{p^--k^- - 
\frac{(p_\perp-k_\perp)^2+m^2-i\epsilon}{2(p^+-k^+)}}
\end{equation}

\begin{equation}
\delta m_3=\frac{ie^2p^+}{2m}\int
\frac{d^2k_\perp}{(2\pi)^4}k^2_\perp\int
\frac{dk^+}{k^{+2}(p^+-k^+)}\int\frac{dk^-}{k^-\bigg[p^--k^-
\frac{(p_\perp-k_\perp)^2+m^2-i\epsilon}{2(p^+-k^+)}\bigg]}
\end{equation}

Using Eq.~(\ref{eq:B.5})-(\ref{eq:B.9}) ,  $\delta m_1$  reduces to
\begin{equation}
\delta m_1  = -\frac{e^2p^+}{2m}\int\frac{d^2k_\perp}{(2\pi)^3}
\int\frac{dk^+}{k^{+2}}[\theta(k^+-p^+)-\theta(p^+-k^+)]
\end{equation}\\
which is the same as $\delta m_c$, 
\begin{equation}
\delta m_2 = \frac{e^2p^+}{2m}\int\frac{d^2k_\perp}{(2\pi)^3}
\bigg[\int_{p^+}^\infty\frac{dk^+}{2k^+(p^+-k^+)}-
\int_{-\infty}^{p^+}\frac{dk^+}{2k^+(p^+-k^+)}\bigg]
\end{equation}\\
and 
\begin{equation}
\delta m_3=\frac{e^2p^+}{2m}\int\frac{d^2k_\perp}{(2\pi)^3}
\bigg[\int^\infty_{p^+}\frac{dk^+}{{k^+}^2}-
\int^{p^+}_{-\infty}\frac{dk^+}{{k^+}^2}\bigg]
\end{equation}
The sum of $\delta m_2$ and $\delta m_3$, on performing $k^+$ 
integration reduces to 
\begin{equation}
\delta m_2+\delta m_3=-\frac{e^2}{2m}\int\frac{d^2k_\perp}{(2\pi)^3}
\bigg[2\bigg(\frac{p^+}{\alpha}-1\bigg)+ln\bigg(\frac{p^+}{\beta}
\bigg)\bigg]
\end{equation}
\\
which is the same as the IR divergent part of $\delta m_a$.  Thus, 
the covariant expression, in the limit  
$k^-\rightarrow \infty, k^+ \rightarrow p^+$, reproduces the sum of 
instantaneous photon exchange graph and the IR singular terms in 
the propagating graph. 

\subsection{\label{subsection:vacp}\bf VACUUM POLARIZATION}

Photon self energy is given by
\begin{equation}
\delta\mu^2\delta_{\lambda\lambda^\prime}
=\epsilon^\lambda_\mu(p)\Pi^{\mu\nu}(p)
\epsilon^{\lambda^\prime}_\nu(p)
\end{equation}\\
where
\begin{equation}
i\Pi^{\mu\nu}=-e^2\int\frac{d^2k_{\perp}}{(2\pi)^4}\int dk^+
\int dk^-\frac{Tr[\gamma^\mu(k \llap/+m)
\gamma^\nu(p \llap/-k \llap/+m)]}{D_1D_2}
\label{eq:vacpol}
\end{equation}

One can rewrite Eq.~(\ref{eq:vacpol})as 
\begin{equation}
i\Pi^{\mu\nu}(p)
=-e^2\int\frac{d^3k}{(2\pi)^3}\int\frac{dk^-}{2\pi}
\frac{Tr[\gamma^\mu(k \llap/+m)\gamma^\nu(p \llap/-k \llap/-m)]}
{2k^+ 2(p^+-k^+) \bigg[k^--\frac{k_\perp^2+m^2-i\epsilon}{2k^+}
\bigg]\bigg[p^--k^--
\frac{(p_\perp-k_\perp)^2+m^2-i\epsilon}{2(p^+-k^+)}\bigg]}
\end{equation}\\
Taking $k^-\rightarrow\infty$ limit in the numerator,we obtain
 
\begin{equation}
\delta \mu_{asy}^2 = ie^2\int\frac{d^2k_\perp}{(2\pi)^4}\int dk^+ 
dk^-\frac{[-4k^+k^-+4(p^+-k^+)k^-+4k_\perp^2]}{D_1D_2}\
\end{equation}
In the limit $k^+\rightarrow0$ and $k^-\rightarrow\infty$ this 
reduces to
\begin{equation}
\delta {\mu^2_ {asy1}}= ie^2\int\frac{d^2k_\perp}{(2\pi)^4}
\int dk^+dk^-\frac{[-4k^+k^-+4(p^+-k^+)k^-+4k_\perp^2]}
{2k^+\bigg(k^--\frac{k^2_\perp+m^2-i\epsilon}{2k^+}\bigg)
(-2)(p^+-k^+)k^-}
\end{equation}
which, on using the Eqs.~(\ref{eq:B.5})-~(\ref{eq:B.9}), reduces to
\begin{eqnarray}
\nonumber
\delta\mu_{asy1}^2&=&\frac{e^2}{2}\int\frac{d^2k_\perp}{(2\pi)^3}
\bigg[\int^\infty_0\frac{dk^+}{(p^+-k^+)}
-\int_{-\infty}^0\frac{dk^+}{(p^+-k^+)}\bigg]\\
\nonumber
&&-\frac{e^2}{2} \int\frac{d^2k_\perp}{(2\pi)^3}\bigg[\int^\infty_0
\frac{dk^+}{k^+}-\int_{-\infty}^0\frac{dk^+}{k^+}\bigg]\\
&&-e^2 \int\frac{d^2k_\perp}{(2\pi)^3}\bigg[\int^0_{\infty}
\frac{dk^+}{p^+-k^+}-\-\int^0_{-\infty}\frac{dk^+}{p^+-k^+}\bigg]
\end{eqnarray}\\

Similarly, in the limit $k^+\rightarrow p^+$, $k^-\rightarrow\infty$,
one obtains
\begin{equation}
\delta\mu_{asy2}^2=ie^2\int\frac{d^2k_\perp}{(2\pi)^4}
\int dk^+dk^-\frac{[-4k^+k^-+4(p^+-k^+)k^-+4k_\perp^2]}
{2k^+k^-2(p^+-k^+)\bigg[p^--k^--
\frac{(p_\perp-k_\perp)^2+m^2+i\epsilon}{2(p^+-k^+)}\bigg]}
\end{equation}

Thus $\delta\mu_{asy2}^2$ becomes
\begin{eqnarray}
\nonumber
\delta\mu_{ asy2}^2&=&\frac{e^2}{2}\int\frac{d^2k_{\perp}}{(2\pi)^3}
\bigg[\int^{\infty}_{p^+}\frac{dk^+}{p^+-k^+}-
\int_{-\infty}^{p^+}\frac{dk^+}{p^+-k^+}\bigg]\\
\nonumber
&&-\frac{e^2}{2}\int\frac{d^2k_{\perp}}{(2\pi)^3}
\bigg[\int^{\infty}_{p^+}\frac{dk^+}{k^+}-\int_{-\infty}^{p^+}
\frac{dk^+}{k^+}\bigg]\\
&&+e^2\int\frac{d^2k_{\perp}}{(2\pi)^3}\bigg[\int^{\infty}_{p^+}
\frac{dk^+}{k^+}-\int_{-\infty}^{p^+}\frac{dk^+}{k^+}\bigg]
\end{eqnarray}
Adding $\delta\mu_{asy1}^2$and $ \delta\mu_{asy2}^2$ we finally 
obtain 
\begin{equation}
\delta\mu_{asy}^2=e^2\int\frac{d^2k_{\perp}} {(2\pi)^3}\bigg[ln
\bigg[\frac{\alpha\beta}{(p^+)^2)}\bigg]\bigg]
\end{equation}
which is the IR singular part of the propagating diagram of one loop 
vacuum polarization in Eq ~(\ref{eq:28}).

\section{\label{sec:sec4} \bf SUMMARY AND CONCLUSION}

We have shown that the instantaneous photon exchange diagrams present 
in one loop fermion self energy calculation within LFTOPT can be 
generated by taking the asymptotic limit  $k^+\rightarrow p^+$,
$k^-\rightarrow\infty$ of the covariant expression. In our earlier 
work \citep{amswa05}, we had shown that the third term in the doubly 
transverse photon propagator is necessary to generate this diagram. 
Here, in this alternative method of generating this diagram, we have 
used the two term photon propagator only. Thus the asymptotic method 
provides an alternative way to generate photon exchange diagrams. In 
addition, this limit also reproduces the IR divergent terms in 
propagating diagrams. This method does not generate instantaneous 
fermion exchange diagrams. It is well established that these diagrams 
arise when one takes the limit  $k^- \rightarrow k^-_{on} $ of 
covariant expression to  obtain propagating diagram of the light front 
perturbation theory. Thus, subtracting the two limits 
$k^- \rightarrow (k^--k^-_{on}) $ and 
$k^- \rightarrow \infty, k^+ \rightarrow p^+ $ will render the 
covariant expression completely free of IR singularities. 

In case of vacuum polarization, there are no instantaneous photon 
exchange diagrams, but the propagating diagram does have an IR 
divergent contribution. In this case, both the internal line are 
fermions and therefore, we consider  both the limits 
$k^- \rightarrow \infty, k^+ \rightarrow 0 $ as well as 
$k^- \rightarrow \infty, k^+ \rightarrow p^+ $ to obtain the IR 
divergent contribution. We verify that the IR singular part of 
propagating term can indeed be generated by this method. Similar to 
the self energy case, one can use this method to subtract the IR 
singular part from the propagating diagrams. It is worth mentioning 
that the IR divergences we have discussed here are not the "true" IR 
divergences of LFFTs \citep{am94,am96}, but are the "spurious" IR 
divergences arising due to the form of LF energy momentum relation. 
True IR divergences shall remain after the above procedure has been 
applied and have to be dealt with separately. 

The procedure sketched here can also be applied to vertex correction 
graphs. We shall address this issue in a future communication.

\begin{acknowledgments}
This work was done under project no.SR/S2/HEP-0017 of 2006 funded by 
Department of Science and Technology in India. A.M. would also like 
to thank Theory Division, CERN,where part of this work was done, for 
their warm hospitality. S.P. would like to thank the Department of 
Physics, University of Mumbai for financial support.
\end{acknowledgments}
\appendix*
\section{A}
\subsection{Basics}
We define the light front co-ordinates by
\begin{equation}
x^+=\frac{x^0+x^3}{\sqrt{2}}
\label{eq:A.1}
\end{equation}
\begin{equation}
x^-=\frac{x^0-x^3}{\sqrt{2}}
\label{eq:A.2}
\end{equation}
\begin{equation}
x_\perp=(x^1,x^2)
\label{eq:A.3}
\end{equation}

  The metric tensor is given by,
\begin{displaymath}
{g^{\mu\nu}}=
\left( \begin{array}{cccc}
{0}&{1}&{0}&{0}\\
{1}&{0}&{0}&{0}\\
{0}&{0}&{-1}&{0}\\
{0}&{0}&{0}&{-1}
\end{array} \right)
\end{displaymath}
 Dirac matrices satisfy the following properties:
\begin{equation}
(\gamma^+)^2=(\gamma^-)^2=0
\label{eq:A.4}
\end{equation}
\begin{equation}
\{\gamma^\mu,\gamma^\nu\}=2g^{\mu\nu}
\label{eq:A.5}
\end{equation}
\begin{equation}
(\gamma^0)^+=\gamma^0 
\label{eq:A.6}
\end{equation}
\begin{equation}
(\gamma^k)^\dagger=-\gamma^k (k=1,2,3)
\label{eq:A.7}
\end{equation}
\begin{equation}
\gamma^+\gamma^-\gamma^+=2\gamma^+
\label{eq:A.8}
\end{equation}
\begin{equation}
\gamma^-\gamma^+\gamma^-=2\gamma^-
\label{eq:A.9}
\end{equation}
\begin{equation}
d_{\mu\nu}(p)= -g_{\mu\nu}+\frac{\delta_{\mu+}p_\nu+
\delta_{\nu+}p_\mu}{p^+}
\end{equation}
also,
\begin{equation}
\gamma^\alpha\gamma^\beta d_{\alpha\beta}(p)=-2
\label{eq:A.10}
\end{equation}
\begin{equation}
\gamma^\alpha\gamma^\nu\gamma^\beta d_{\alpha\beta}(p)=\frac{2}{p^+}
(\gamma^+\gamma^\nu+g^{+\nu} p\llap/)
\label{eq:A.11}
\end{equation}
\begin{equation}
\gamma^\alpha\gamma^\mu\gamma^\nu\gamma^\beta d_{\alpha\beta}(p) = 
-4g^{\mu\nu}+2\frac{p_\alpha}{p^+}(g^{\mu\alpha}\gamma^\nu\gamma^+-
g^{\alpha\nu}\gamma^\mu\gamma^++g^{\alpha+}\gamma^\mu\gamma^\nu-
g^{+\nu}\gamma^\mu\gamma^\alpha+g^{+\mu}\gamma^\nu \gamma^\alpha)
\label{eq:A.12}
\end{equation}

Dirac spinors satisfy:
\begin{equation}
\bar u(p,s)u(p,s^\prime) = -\bar v(p,s)v(p,s) = 2m\delta_{s,s^\prime}
\label{eq:A.12}
\end{equation}

\begin{equation}
\bar u(p,s)\gamma^\mu u(p,s^\prime) = \bar v(p,s)\gamma^\mu v(p,s) 
= 2p^\mu \delta_{s,s^\prime}
\label{eq:A.13}
\end{equation}

\subsection{Light Front Hamiltonian}

$P^-$, the Light front  Hamiltonian is  the operator conjugate to the 
``time'' evolution variable  $x^+$  and is given by,
\begin{equation}
P^-=H_0+V_1+V_2+V_3
\label{eq:A.14}
\end{equation}
where $H_0$ is the free hamiltonian,
$V_1$ is the standard, order-e three-point interaction,
\begin{equation}
V_1=e\int d^2x_\perp dx^- \bar\xi \gamma^\mu \xi a_\mu
\label{eq:A.15}
\end{equation}
$V_2$ is an order-$e^2$ non-local effective four- point vertex 
corresponding to an instantaneous fermion exchange, 
\begin{equation}
V_2=-\frac{i}{4}e^2\int d^2x_\perp dx^- dy^-\epsilon(x^- -y^-)
(\bar\xi a_k \gamma^k)(x) \gamma^+(a_j \gamma^j\xi)(y)
\label{eq:A.16}
\end{equation}
and $V_3$ is an order-$e^2$ non-local effective four-point vertex 
corresponding to an instantaneous photon exchange, 
\begin{equation}
V_3
=-\frac{e^2}{4}\int d^2x_\perp dx^-dy^-(\bar\xi \gamma^+\xi)(x)
|x^--y^-| (\bar\xi\gamma^+\xi)(y)
\label{eq:A.17}
\end{equation}
\subsection{Instantaneous diagrams in Self Energy correction}

Here, we briefly review the calculation of $\delta m_b$ and $\delta m_c$ in 
Eqs.~(\ref{eq:mb}) and ~(\ref{eq:mc}). The details can be found in Appendix B of Ref.~\cite{wilson91}. 
To prove the expression for $\delta m_b$ in Eq.~(\ref{eq:mb}) starting from 
Eq.~(\ref{eq:m2}), one writes 
\begin{eqnarray}
\nonumber
\int_0^\infty dk^+\frac{p^+}{k^+(p^+-k^+)} = \int_0^\infty dk^+\bigg[\frac{1}{k^+} + \frac{1}{p^+-k^+}\bigg]\\
 = \int_\alpha^\infty\frac{dk^+}{k^+}+\int_{p^++\eta}^\infty\frac{dk^+}{p^+-k^+}+\int_0^{p^+-\eta}\frac{dk^+}{p^+-k^+}= ln\bigg[\frac{p^+}{\alpha}\bigg]
\end{eqnarray}
where we have identified $\alpha$ with $\eta$.

To prove Eq.~(\ref{eq:mc}), we start with Eq.~(\ref{eq:m3}) and write
\begin{eqnarray}
\nonumber
\int_0^\infty \frac{dk^+}{(p^+-k^+)^2} -\int_0^\infty\frac{dk^+}{(p^++k^+)^2}=
\int_0^{p^+-\eta}\frac{dk^+}{(p^+-k^+)^2}\\
\nonumber
+\int_{p^++\eta}^\infty\frac{dk^+}{(p^+-k^+)^2}-\int_{p^+}^\infty\frac{dk^+}{(k^+)^2}\\
\nonumber
 = \bigg[\int_\eta^\infty+\int_\eta^{p^+}-\int_{p^+}^\infty\bigg]\frac{dk^+}{(k^+)^2}\\
=2\int_\eta^{p^+}\frac{dk^+}{(k^+)^2}=\frac{2}{p^+}\bigg[\frac{p^+}{\alpha}-1\bigg]
\end{eqnarray}
where again we have identified $\eta $ with $\alpha$.

\appendix*
\section{B}
\setcounter{equation}{0}
\renewcommand{\theequation}{B.\arabic{equation}}
In this appendix ,we will give expressions for the integrals used 
in Section ~\ref{sec:sec3}. Consider the integral 
\begin{equation}
I _1 = \int^\infty_0\frac{dk^-}{k^--\frac{k^2_\perp+\mu^2-i\epsilon}
{2k^+}}
\end{equation}
which has a pole at $k^-=\frac{k^2_\perp+\mu^2-i\epsilon}{2k^+}$ 
that tends to infinity in the limit $k^+\rightarrow 0$. To evaluate 
the integral, we change the variable to $u = \frac{1}{k^-}$ and 
obtain 

\begin{equation}
I _1=\int\frac{du}{u\bigg[1-
\frac{k^2_\perp+\mu^2-i\epsilon}{2k^+}u\bigg]}
\end{equation}

Regularizing the integral by the replacement 
\begin{equation}
\frac{1}{u} = \frac{1}{2}\bigg(\frac{1}{u+i\delta}+
\frac{1}{u-i\delta}\bigg)
\end{equation}
we obtain 
\begin{equation}
I_1 =\frac{1}{2}\int\frac{du}{(u+i\delta)\bigg[1-
\frac{k^2_\perp+\mu^2-i\epsilon}{2k^+}u\bigg]}+\frac{1}{2}
\int\frac{du}{(u-i\delta)\bigg[1-\frac{k^2_\perp+
\mu^2-i\epsilon}{2k^+}u\bigg]}
\end{equation}
Closing the contour in the lower half plane for the first integral 
and in the upper half plane for the second integral, we finally 
obtain
\begin{equation}
I_1 = -\pi i[\theta(k^+)-\theta(-k^+)]
\label{eq:B.5}
\end{equation}

Similarly the integral 
\begin{equation}
I_2 = \int\frac{dk^-}{p^--k^--
\frac{(p_\perp-k_\perp)^2+m^2-i\epsilon}{2(p^+-k^+)}}
\end{equation}
has a pole at $k^- = p^--\frac{(p_\perp-k_\perp)^2+m^2-i\epsilon}
{2(p^+-k^+)}$ which tends to infinity as $k^+\rightarrow p^+$. Again 
changing the variable to $u= \frac{1}{k^-}$ and using the same 
procedure as above we finally obtain 
\begin{equation}
\int\frac{dk^-}{p^--k^--\frac{(p_\perp-k_\perp)^2+m^2-i\epsilon}
{2(p^+-k^+)}}=\pi i[\theta(k^+-p^+)-\theta(p^+-k^+)]
\label{eq:B.7}
\end{equation}
Also the integral
\begin{equation}
\int\frac{dk^-}{k^-\bigg[k^--\frac{k_\perp^2+\mu^2-i\epsilon}{2k^+}
\bigg]}=-\pi i  \frac{2k^+  [\theta(k^+)-\theta(-k^+)]}{k_\perp^2+\mu^2
-i\epsilon}
\end{equation}
and
\begin{equation}
\int\frac{dk^-}{k^-\bigg[p^--k^--
\frac{(p_\perp-k_\perp)^2+m^2-i\epsilon}{2(p^+-k^+)}\bigg]}=\pi i 
\frac{2(p^+-k^+)[\theta(k^+-p^+)-\theta(p^+-k^+)]}{2(p^+-k^+)p^--
(p_\perp-k_\perp)^2-m^2+i\epsilon}
\label{eq:B.9}
\end{equation}

\end{document}